\documentclass{aa} 
\usepackage{natbib}
\bibpunct{(}{)}{;}{a}{}{,} 

\usepackage[colorlinks=true,linkcolor=blue,allcolors=blue]{hyperref}
\usepackage[flushleft]{threeparttable}
\usepackage{tabularx}

\usepackage{tabularx}
\usepackage{amsmath}
\usepackage{amsfonts}
\usepackage{amssymb}
\usepackage{mathtools}
\DeclarePairedDelimiter{\ceil}{\lceil}{\rceil}

\usepackage{graphicx}
\usepackage{txfonts}
\begin{document}

   \title{An examination of the effect of the TESS extended mission on southern hemisphere monotransits}

   \author{Benjamin F. Cooke,\thanks{b.cooke@warwick.ac.uk}\inst{\ref{inst1},\ref{inst2}}
        Don Pollacco\inst{\ref{inst1},\ref{inst2}}
        \and
        Daniel Bayliss\inst{\ref{inst1},\ref{inst2}}
        }

        \institute{
        Department of Physics, University of Warwick, Gibbet Hill Road, Coventry CV4 7AL, UK\\
        \label{inst1}
        \and
        Centre for Exoplanets and Habitability, University of Warwick, Gibbet Hill Road, Coventry CV4 7AL, UK\\
        \label{inst2}
        }

   \date{Received September 15, 1996; accepted March 16, 1997}

 
  \abstract
   {
   NASA recently announced an extended mission for TESS. As a result it is expected that the southern ecliptic hemisphere will be re-observed approximately two years after the initial survey.
   }
   {We aim to explore how TESS re-observing the southern ecliptic hemisphere will impact the number and distribution of monotransits discovered during the first year of observations. This simulation will be able to be scaled to any future TESS re-observations.}
   {We carry out an updated simulation of TESS detections in the southern ecliptic hemisphere. This simulation includes realistic Sector window-functions based on the first 11 sectors of SPOC 2\,min SAP lightcurves. We then extend this simulation to cover the expected Year 4 of the mission when TESS will re-observe the southern ecliptic fields. For recovered monotransits we also look at the possibility of predicting the period based on the coverage in the TESS data.}
   {We find an updated prediction of 339 monotransits from the TESS Year 1 southern ecliptic hemisphere, and that approximately 80\% of these systems (266/339) will transit again in the Year 4 observations. The Year 4 observations will also contribute new monotransits not seen in Year 1, resulting in a total of 149 monotransits from the combined Year 1 and Year 4 data sets.  We find that 75\% (189/266) of recovered Year 1 monotransits will only transit once in the Year 4 data set.  For these systems we will be able to constrain possible periods, but period aliasing due to the large time gap between Year 1 and Year 4 observations means that the true period will remain unknown without further spectroscopic or photometric follow-up.}
   {}


    \titlerunning{TESS Year 4}
    \authorrunning{Benjamin F. Cooke et al.}

   \maketitle

\section{Introduction}
\label{sec:Introduction}

The Transiting Exoplanet Survey Satellite \citep[\textit{TESS},][]{Ricker2015} has recently completed its first year of observations of the southern ecliptic hemisphere. It will now move its observations to the north, continuing its full-sky survey. At the time of writing TESS has already successfully discovered and confirmed 29 new transiting planets \citep[NASA Exoplanet Archive\footnote{\href{https://exoplanetarchive.ipac.caltech.edu/index.html}{https://exoplanetarchive.ipac.caltech.edu/index.html}},][]{Akeson2013} including $\pi$ Mensae\,c \citep{huang2018b,Gandolfi2018}, HD\,202772A\,b, \citep{Wang2019}, LHS\,3844\,b \citep{Vanderspek2019} and HD\,219666\,b, \citep{Esposito2019}.

Due to the all-sky nature of TESS, as well as its relatively short baselines for much of the sky, it produces a large yield of mono-transiting systems: planets which will exhibit only a single transit during TESS observations. The predicted yield of these monotransits has been estimated in \cite{Cooke2018} and \cite{Villanueva2018}. Recently it has been announced that the TESS mission will be extended, and as part of the extension plans it is envisaged that TESS will re-observe the southern ecliptic hemisphere in the fourth year of its mission 
(see NASA 2019 senior review\footnote{\href{https://smd-prod.s3.amazonaws.com/science-pink/s3fs-public/atoms/files/SR2019_Subcommittee_Rpt.pdf}{2019 Senior Review Subcommittee Report}} and NASA response\footnote{\href{https://smd-prod.s3.amazonaws.com/science-pink/s3fs-public/atoms/files/NASA_Response_SR2019_Final.pdf}{NASA Response to the 2019 Senior Review of Operating Missions}}). Theoretical extended mission templates have been suggested by \cite{Huang2018} and \cite{Bouma2017}. At the time of writing some of the specifics of this extended mission are unknown, but it is known that after finishing Year 2 (northern ecliptic) TESS will spend approximately one year observing the ecliptic equator before returning to the south and repeating its Year 1 observations. This is very similar to the suggested `hemi' scenarios presented in both \cite{Huang2018} and \cite{Bouma2017} except we now include an additional intermediary year. An extra detail now currently known is that during the extended mission Postage Stamp (PS) cadence will remain at 2\,mins but Full-Frame Images (FFIs) will have an improved cadence of 10\,mins (c.f. 30\,mins for Year 1 and Year 2 observations).

This extended mission will effect all areas of TESS science, but perhaps most noticeably, the yield of monotransits. Therefore, an extended simulation is required to explore these effects. Since only the return to the southern ecliptic is confirmed we constrain our simulation to this hemisphere under the following assumptions:
\begin{itemize}
    \item In Year 4, TESS will observe each southern ecliptic sector exactly three years after the sector was observed in Year 1.
    \item Each Year 4 sector will match up exactly with the Year 1 sectors.
    \item PS cadence will remain at 2\,min, but FFI cadence will be improved to 10\,min.
    \item Photometric data quality and time coverage will remain constant between Years 1 and 4.
\end{itemize}

This letter is set out in the following way. Section \ref{sec:Methodology} outlines the simulation, particularly focusing on the improvements made to that presented in \cite{Cooke2018}. Section \ref{sec:Results} outlines the key results, first giving an update to the \cite{Cooke2018} results then moving on to the Year 4 results. Finally, Section \ref{sec:Conclusions} sums up the key findings and outcomes.

\section{Methodology}
\label{sec:Methodology}

\subsection{Stellar population}
\label{sec:Stellar population}

We set up our simulation using a similar method as discussed in \cite{Cooke2018}. We begin with our stellar catalogue, the TESS Input Catalogue (TIC) Candidate Target List (CTL) available from the Mikulski Archive for Space Telescopes (MAST\footnote{\href{http://archive.stsci.edu/tess/tic\_ctl.html}{http://archive.stsci.edu/tess/tic\_ctl.html}}). This simulation employs the latest version, version 8, of the CTL cross-matched against the TIC as described by \citet{Stassun2019}. 
We then filter this sample using TESS-band magnitude cuts, $3.0 \leq m_{TESS} \leq 17.0$, and effective temperature cuts, $2285 \leq T_{eff} \leq 10050$. 
Taking only those stars in the southern ecliptic hemisphere produces a final input stellar catalogue of 4789372 potential hosts.

For each target we then predict how many TESS sectors will observe it. We simulate 13 sectors, each $24^\circ\times96^\circ$ on the sky, and, using ecliptic coordinates, find the number of sectors overlapping each star. 

\subsection{Planetary population}
\label{sec:Planetary population}

With the stellar catalogue known we then simulate a planetary population around the stars in the same way as discussed in \cite{Cooke2018}. Occurrence rates as functions of radius and period are taken from \cite{Dressing2015} and \cite{Fressin2013} depending on spectral type and the hosts are randomly populated with planetary companions. 
%
%
%
%
%
%
Transit parameters such as depth and duration are calculated following the formalism set out in \cite{Cooke2018}, which use the equations from \cite{Winn2014} and \cite{Barclay2018}.

\subsection{Detectability}
\label{sec:Detectability}

S/N, and thus detectability, is a function of transit depth, stellar contamination, instrumental noise, and number of data points in transit. We calculate this value using the equations presented in \cite{Cooke2018}, employing the $5^{\rm th}$ order polynomial noise approximation given in \cite{Stassun2017}.








Lastly, we determine host cadence using a priority metric,

\begin{equation}
\label{eq:metric}
\frac{\sqrt{N_s}}{\sigma_{1hr} R_{\star}^{3/2}},
\end{equation}

where $N_s$ is the number of sectors for which a target is observed, $\sigma_{1hr}$ is the photometric noise in an hour and $R_{\star}$ is stellar radius. The top ranked 200,000 stars are given an observing cadence of 2\,min, corresponding to the Postage Stamp (PS) targets, with the rest getting 30\,min cadence during the first year of observing and 10\,min cadence when TESS returns to the south hemisphere in year 4 corresponding to the Full-Frame Images (FFIs).

Although it is now known exactly which stars in the TIC have been observed at 2\,min cadence in Year 1 the metric used here should produce an almost identical population. Since we do not yet know the 2\,min targets for the extended mission it was decided to continue with the metric for both the primary and extended mission.

\subsection{Sector Window Functions}
\label{sec:Sector masks}

To make this simulation more realistic we create and use sector window functions from the TESS data currently available for the southern sectors. In the simulation presented in \cite{Cooke2018} we assumed that each sector was two 13.7 day orbits with 6.5hrs of downtime per orbit. In reality the true amount of usable observing time is less than this due to a variety of reasons including momentum dumps, high scattered light, and spacecraft/camera down-times. To take this into account we create window functions for each sector and assume that only data points falling within these window functions are reliable observations.

To create these window functions we examine the publicly available 2\,min lightcurves from the TESS Science Processing Operations Center pipeline \citep[SPOC,][]{Jenkins2016}. We download 16 representative lightcurves from each sector, one each from each of the four CCDs on each of the four cameras. For these lightcurves we then compare the Simple Aperture Photometry (SAP) and Pre-search Data Conditioned (PDC\_SAP) lightcurves. At each timestamp we compare the two lightcurves, if a timestamp has not been translated over from SAP to PDC\_SAP we assume this is due to some instrumental data issue and therefore that data point is unreliable. 
From this we create a window function of good timestamps, those points which have a corresponding PDC\_SAP data point. This results in 16 window functions per sector. Comparisons of these window functions show that differences between CCDs and even between cameras are negligible within a single sector. Thus we use the camera 1, CCD 1 window function as the representative window function for that sector. Since the data for sectors 12 and 13 are not yet publicly available we copy the mask from sector 11 into the sector 12 and 13 time slots. The coverage fraction for the TESS sectors ranges from 98\% (in Sector 2) down to 69\% (in Sector 3), with a mean coverage of 91\% over the first 11 sectors.

Even during this first year there have been changes in data collection from Sector 1 through to Sector 11. However it is hard to estimate how these improvements will translate into the extended mission, as the spacecraft/cameras will aged, and this may effect data quality.  To produce a baseline study we therefore assume the sector window functions created from the first years data can be exactly mapped into Year 4 by simply increasing the timestamps of the window functions by 1095.75 days (3 years).

These sector window functions then help inform the simulation as to how many transits, and crucially, how many data points in transit, will be observed during the TESS Year 1 and Year 4 observations of the southern ecliptic hemisphere. 

For 2\,min cadence observations we take the full window function of good timestamps. For the FFI images in Year 1 and Year 4 (30 and 10\,min cadence respectively) we take a slightly more involved approach. We assume that, if the data collection were perfect, we would take an FFI data point every 30 or 10 minutes for year 1 and 4. Thus the range of FFI timestamps is found by taking every 15th and 5th timestamp from the full SAP 2\,min lightcurves for 30 and 10\,min cadence respectively. For each of these times we cross-reference with the window function of good data. For 30\,min cadence we require that each simulated timestamp has at least 14 good data points in the 30 minutes surrounding it. Any less and we assume the lack of quality data compromises the point. For the 10\,min cadence we require 5 good points in the 10 minutes surrounding each simulated data point. Where this criteria is met we allow the simulated timestamp, where it is not we assume that the timestamp is lost. Thus for each cadence value, 2, 10 and 30\,min, we are left with window functions that we use for the simulation.

Next we simulate the full range of transit times for each planet. We initialise the first transit of a planet at time $T_0$ where $T_0$ is the start of Year 1, Sector 1 observations plus a random fraction of the orbital period. Additional transit times are then simulated up to the end of Sector 13 observations in Year 4. We next use these transit times to reduce our lists of timestamps, keeping only those that land within $T_{dur}/2$ of a transit centre. This process is repeated twice, once using the 2\,min list for Years 1 and 4 and once using the 30\,min list for Year 1 and the 10\,min list for year 4.


This results in a list of $N_{in}$ timestamps per planet where $N_{in}$ is the number of in-transit observations (accounting for observing cadence), spread across $n_{tr}$ transits where $n_{tr}$ is the number of unique transit times that have surviving timestamps within $T_{dur}/2$. This value of $N_{in}$ is then used 
to calculate $S/N$. Only planets with $S/N \geq 7.3$ are classified as being detected.

We keep the Year 1 and Year 4 results separate. Year 4 observations have no bearing on the S/N, and therefore detectability, of planets observed in Year 1 and vice versa. A Year 1 monotransit is any planet with $n_{tr}=1$ and $S/N \geq 7.3$ based on Year 1 data alone and the same definition is used for Year 4 monotransits from Year 4 data.

\section{Results}
\label{sec:Results}

\subsection{Year 1}
\label{sec:Year 1}

Using this simulation we first update the results of \cite{Cooke2018} taking into account the more realistic sector window functions and the updated TIC CTL. As the key results we reproduce Tables 2 and 3 from \cite{Cooke2018} (see Appendix \ref{sec:Year 1 updated results}).

The total planet numbers reported here are noticeably more than expected if one simply halves the values shown in Table 2 of \cite{Cooke2018}. This is in part due to the use of an updated stellar input catalogue and therefore a larger number of simulated systems. Additionally we see more short period monotransits due to the gaps present in the actual TESS window functions.


\subsection{Year 4}
\label{sec:Year 4}

The key issue we seek to address in this work is how an extended mission will affect the population of TESS monotransits. 
We therefore compare the Year 1 and Year 4 results. Note that both years analyse the same population of stars and planets. 
The most important result from this analysis is the fraction of simulated Year 1 monotransits that will have a subsequent transit observed in the Year 4 data.  We find that $\sim$80\% (266 of 339) of the Year 1 monotransits will exhibit at least one detectable transit during Year 4. The remaining 20\% of systems (73 Year 1 monotransits) will remain as single transit systems even after the extended mission. 

\begin{figure}[ht]
\includegraphics[width=\columnwidth]{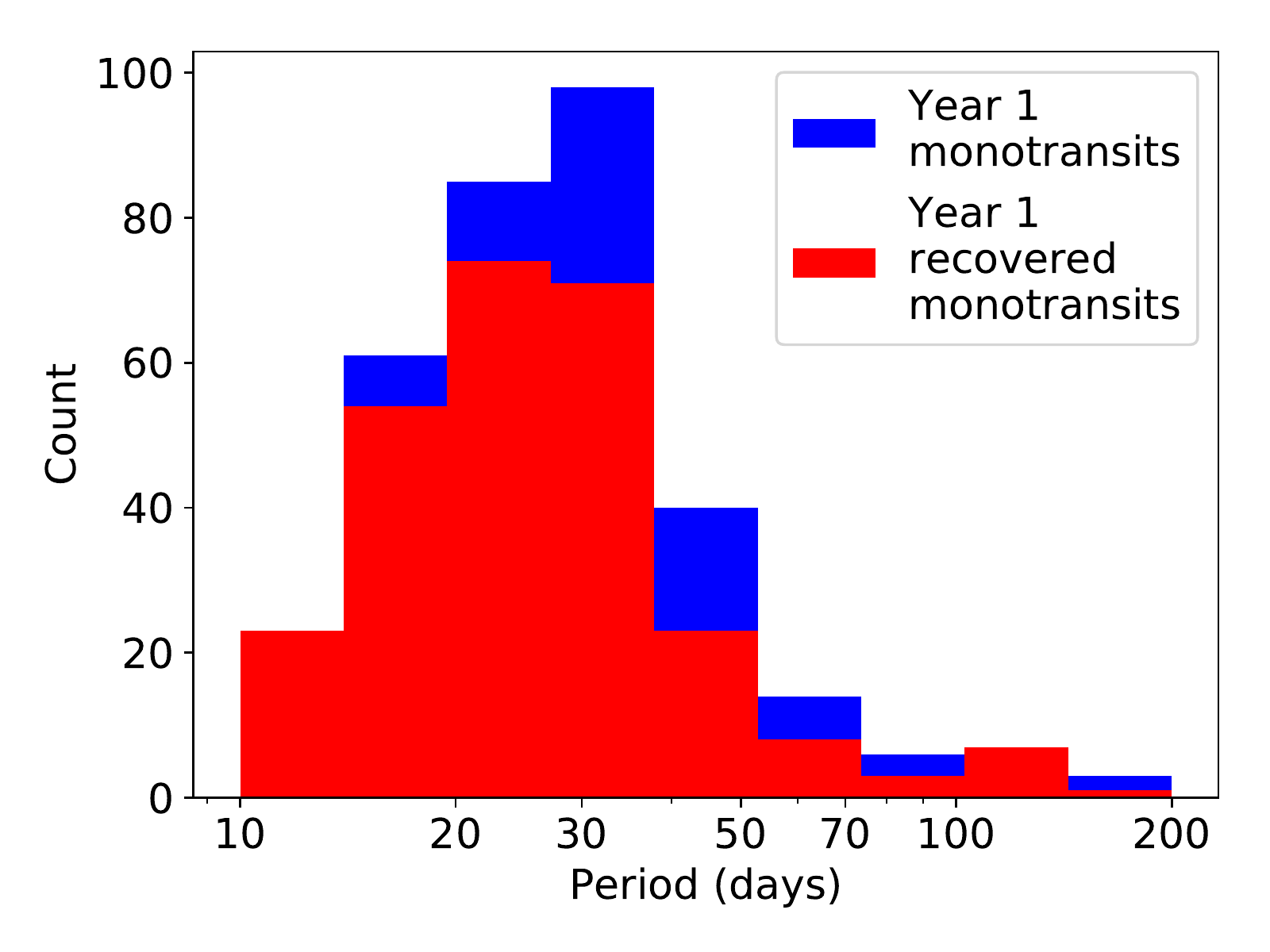}
\caption{The period distribution of Year 1 detectable monotransits (blue) and the fraction that have a recovered transit in the Year 4 data (red).  As expected the shorter period systems show a higher fraction Year 4 recovered transits.}
\label{fig:year1_period}
\end{figure}

Figure \ref{fig:year1_period} show a histogram of Year 1 monotransits as a function of period including those that will transit again in Year 4.  As expected, it is evident that the short period Year 1 monotransits (P$<$20\,d) are largely recovered again in the while the longer period transits (P$>$40\,d) are only recovered in about half the cases. 

In addition to the monotransits from Year 1, the Year 4 data also contains a population of monotransits that did not transit in the Year 1 data.  This population is approximately equal in size to the number of year 1 monotransits that are not found again in Year 4 meaning that the total number of mono-transiting systems after Year 4 is 149 with approximately half being observed to transit in Year 1 and half in Year 4. 

\begin{figure}[ht]
\includegraphics[width=\columnwidth]{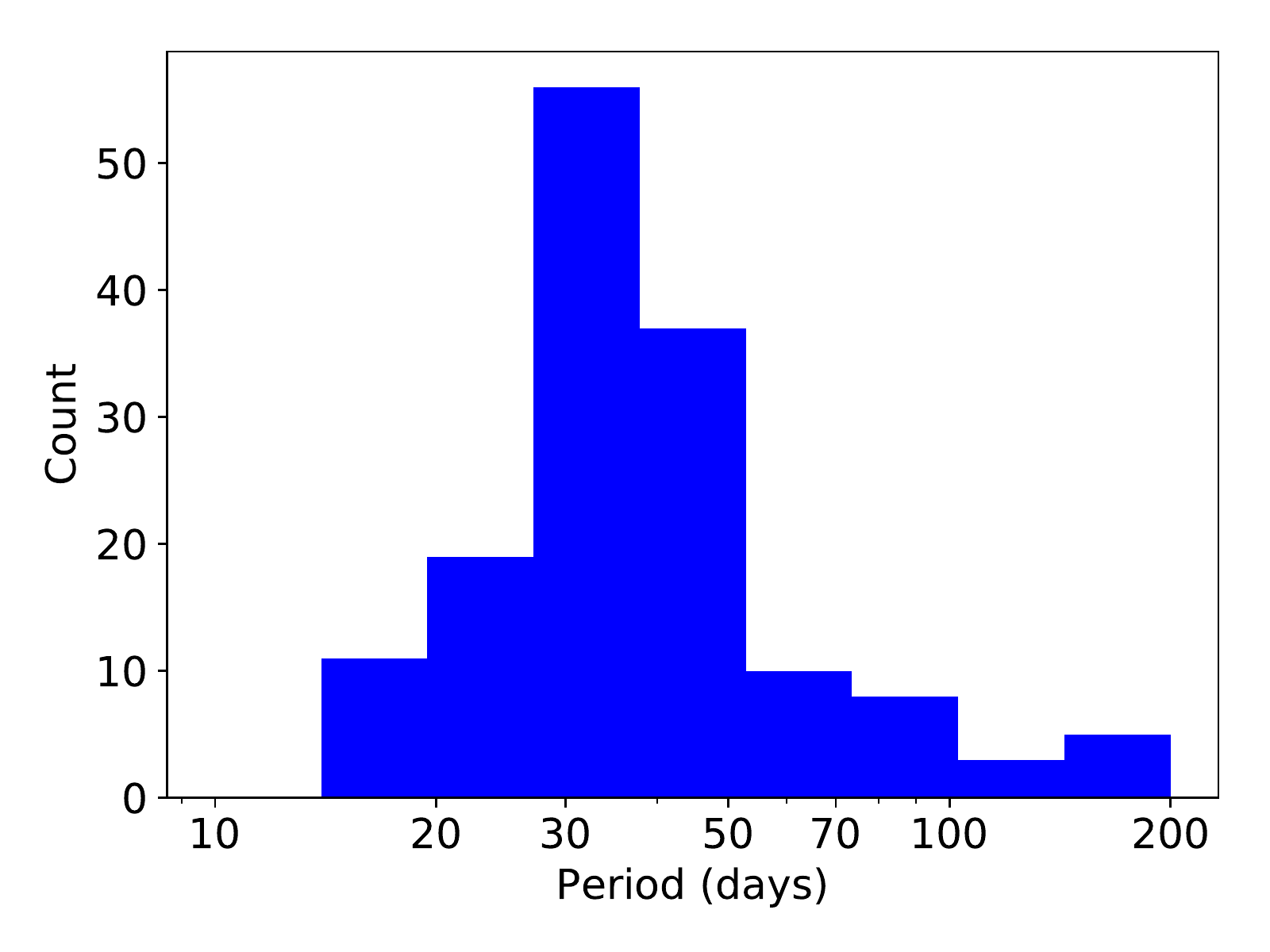}
\caption{Period distribution of systems that remain as monotransits once the Year 1 and Year 4 data have been obtained. Approximately half of these systems are found in the Year 1 data, and half are in the Year 4 data. Note there are far fewer than the original monotransit yield from the Year 1 alone, and the distribution is shifted towards longer period systems.}
\label{fig:year1_4_period}
\end{figure}

Figure \ref{fig:year1_4_period} shows the period distribution of monotransiting planets after Year 4. Comparing this with the distribution of Year 1 monotransits shown in Fig. \ref{fig:year1_period} shows that this population is smaller and distributed more to longer periods as expected.  


\begin{figure}[ht]
\includegraphics[width=\columnwidth]{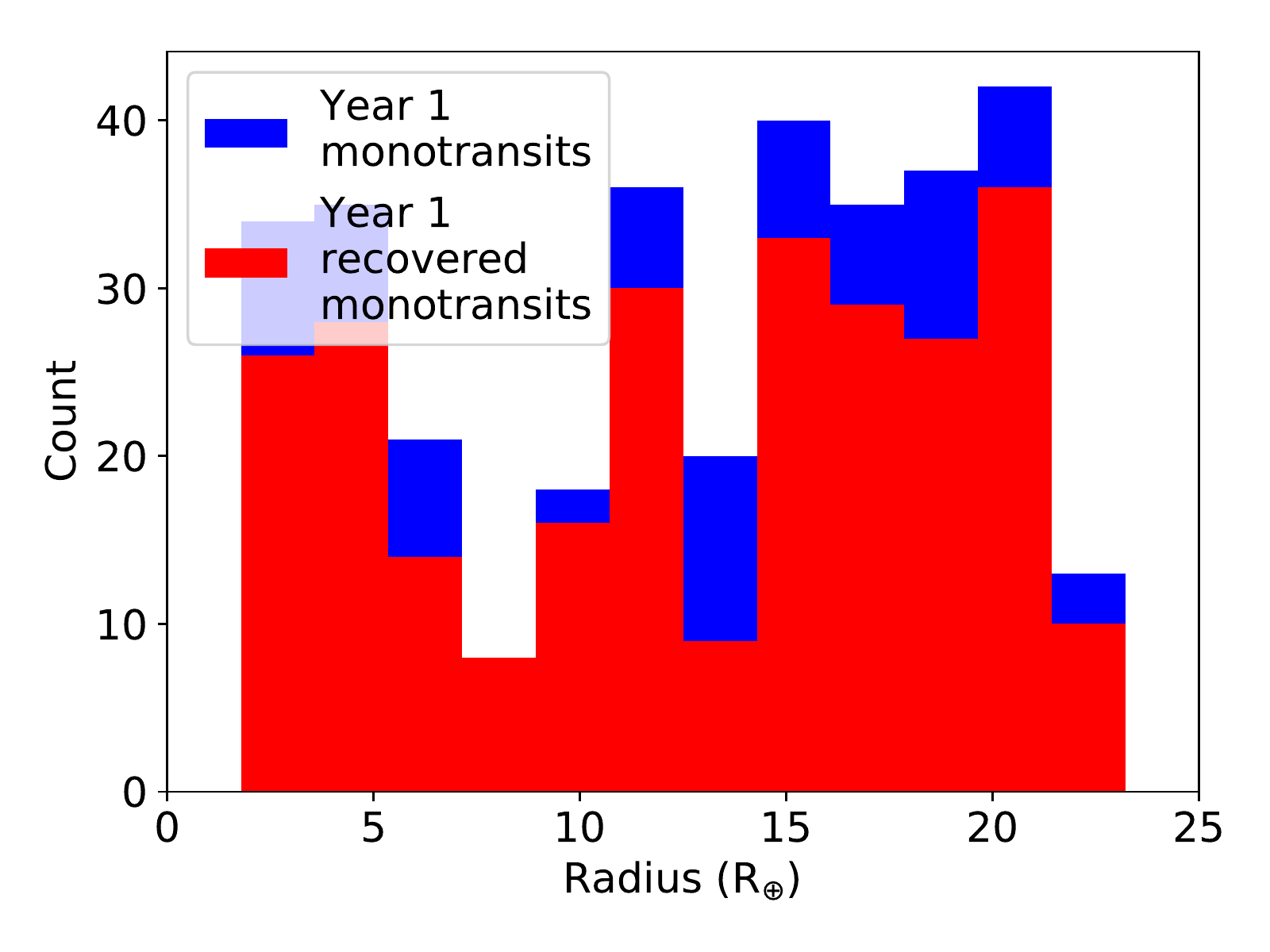}
\caption{The planetary radius distribution of Year 1 detectable monotransits (blue) and the fraction that have a recovered transit in the Year 4 data (red).}
\label{fig:radius}
\end{figure}

\begin{figure}[ht]
\includegraphics[width=\columnwidth]{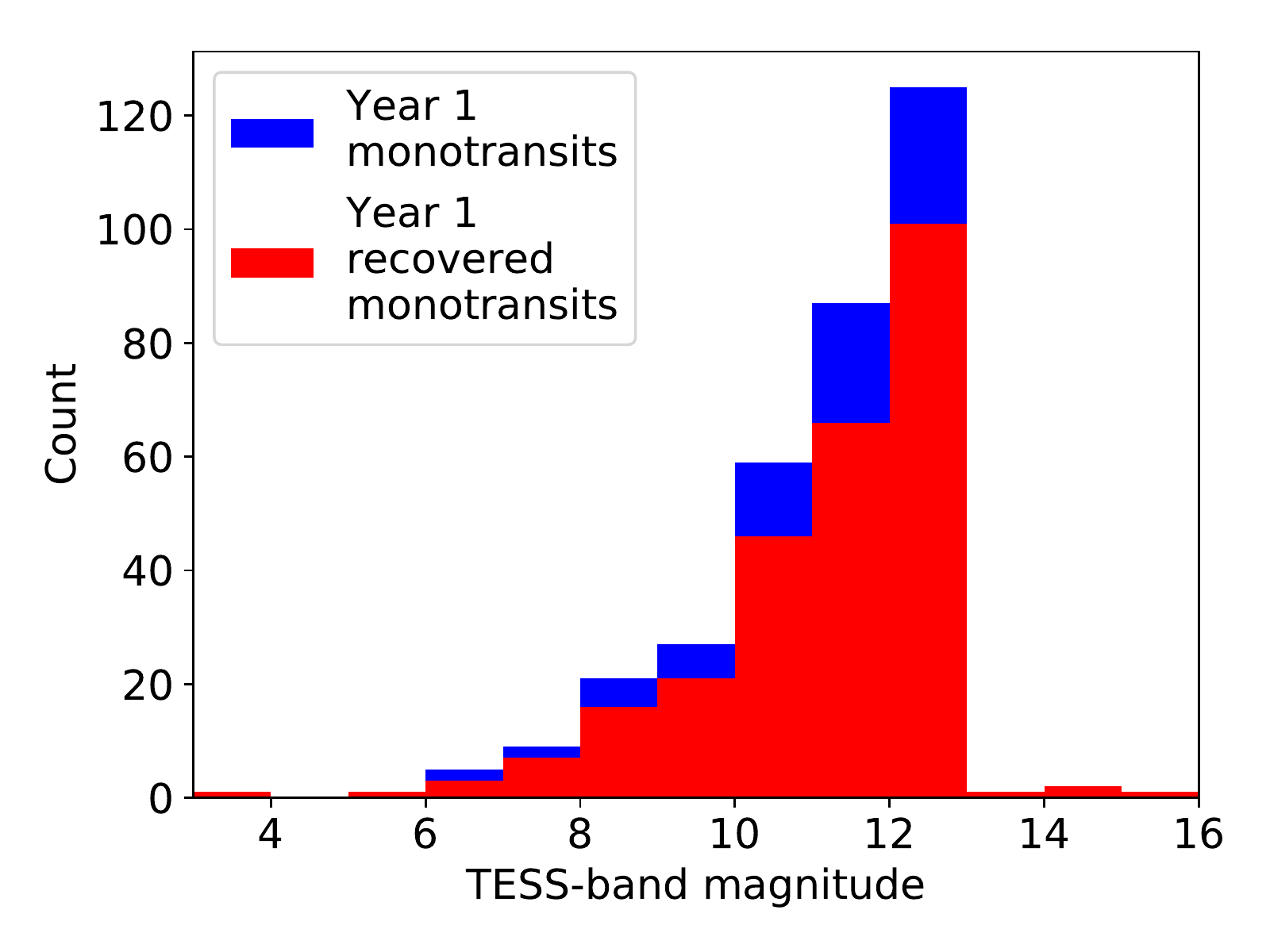}
\caption{The stellar magnitude distribution of Year 1 detectable monotransits (blue) and the fraction that have a recovered transit in the Year 4 data (red).}
\label{fig:mag}
\end{figure}

\begin{figure}[ht]
\includegraphics[width=\columnwidth]{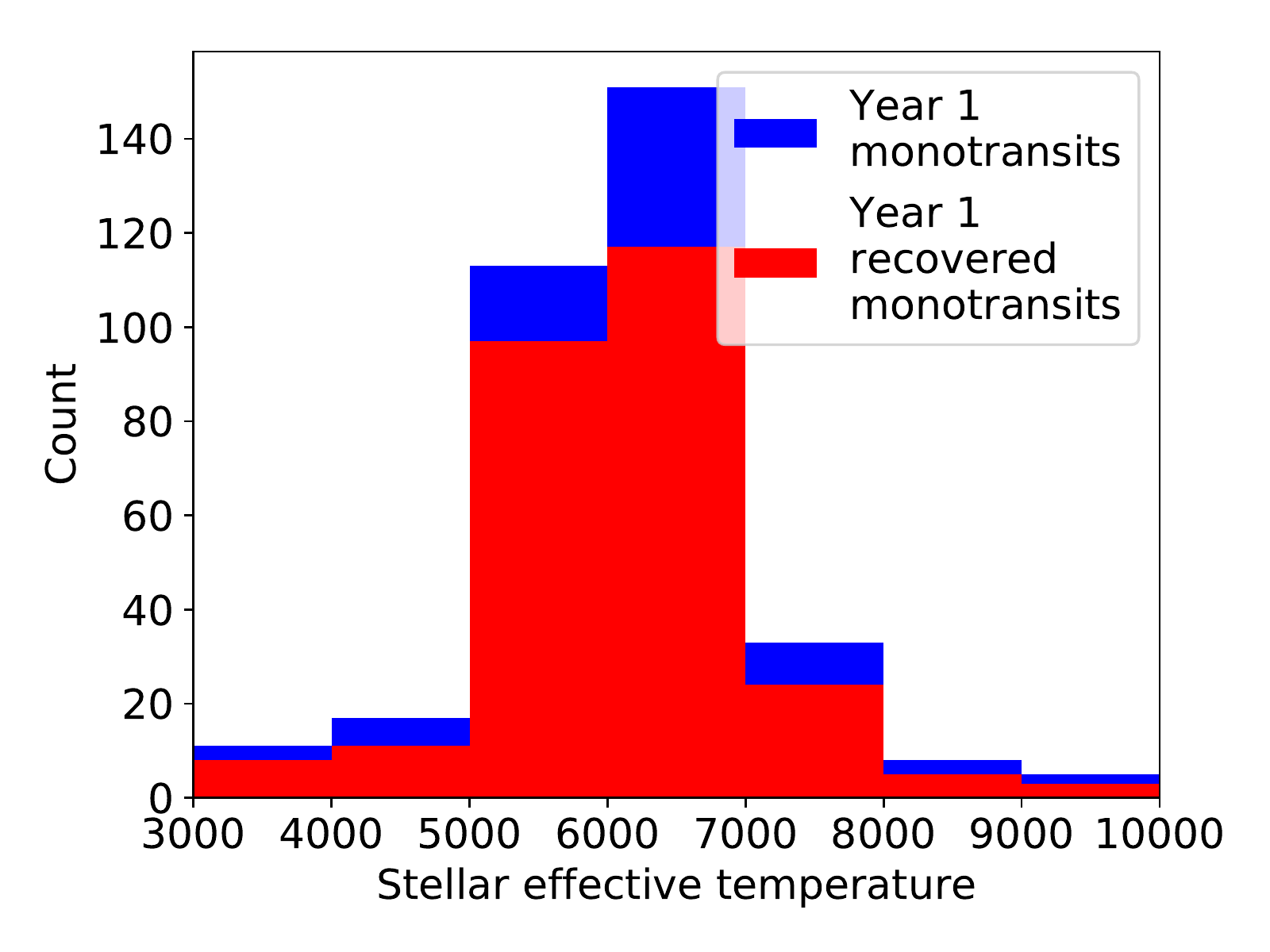}
\caption{The stellar effective temperature distribution of Year 1 detectable monotransits (blue) and the fraction that have a recovered transit in the Year 4 data (red).}
\label{fig:teff}
\end{figure}

It is also interesting to look at how the population of recovered Year 1 monotransits varies with other system parameters. Figures \ref{fig:radius}, \ref{fig:mag} and \ref{fig:teff} show the Year 1 monotransits as well as those observed to transit again in Year 4 as functions of planetary radius, stellar magnitude and stellar effective temperature respectively.

These three figures suggest that the recovered population of Year 1 monotransits is broadly independent of planetary radius, stellar magnitude or stellar effective temperature. Logically this make sense. Targets will be re-observed for the same amount of time in Year 4 as in Year 1, Thus, if detected in Year 1, should be detected again in Year 4 assuming a transit occurs again. Therefore the only meaningful parameter for recovery should be orbital period.

\subsection{Period alias}
\label{sec:Period alias}

Although $\sim$80\% (266 of 339) of Year 1 monotransits will been seen to transit in Year 4 (Section~\ref{sec:Year 4}), approximately 75\% of these (198/266) will only transit \textit{once} in the Year 4 data.  For these systems we will have only two transits, which means rather than being able to determine a single period, there will be a discrete set of allowed periods. 
If we assume no significant transit timing variations, the period of these systems is given by $P = T_{diff}/n$ where $T_{diff}$ is the difference between the Year 1 and Year 4 transit times and $n$ is an integer. For this simulation the average value of $T_{diff}$ is 3 years.


 
Obviously, for real data sets the true value of $n$, and thus $P$, is unknown so next we look at the possibility of using the TESS data to constrain these values. To begin we assume that the smallest possible period for one of these systems is 10 days, any shorter than this and a second transit would have been seen within a single sector, even allowing for gaps in coverage. From this we then create a list of $n$ values for each system with $n$ running from 1 to $n_{max}$ where $n_{max} = \ceil[\Big]{T_{diff}/10}$. For each value of $n$ we generate a list of periods and the corresponding times at which the system would transit, including values before and after the observed transits. These lists are then compared against the sector window functions (allowing for which sectors observe each target) and if any transits fall into regions that have TESS coverage (in Year 1 or 4) we rule out that value of $n$ since a third transit would have been seen. In this way we generate a list of allowed $n$ values (and hence periods) for each system that we can compare against the true period.  In order to compare systems with different periods, we define the true period index $I$ to be

\begin{equation}
\label{eq:I}
I = n_{max} - n_{true},
\end{equation}
where $n_{true}$ is the actual $n$ value for a given system. Thus if the true period is the shortest possible period, the period index will be $I=0$.  A system with a period index of $I=5$ would, for example, mean that the true period of this system is the $6^{\rm th}$ shortest possible period.

\begin{figure}[ht]
\includegraphics[width=\columnwidth]{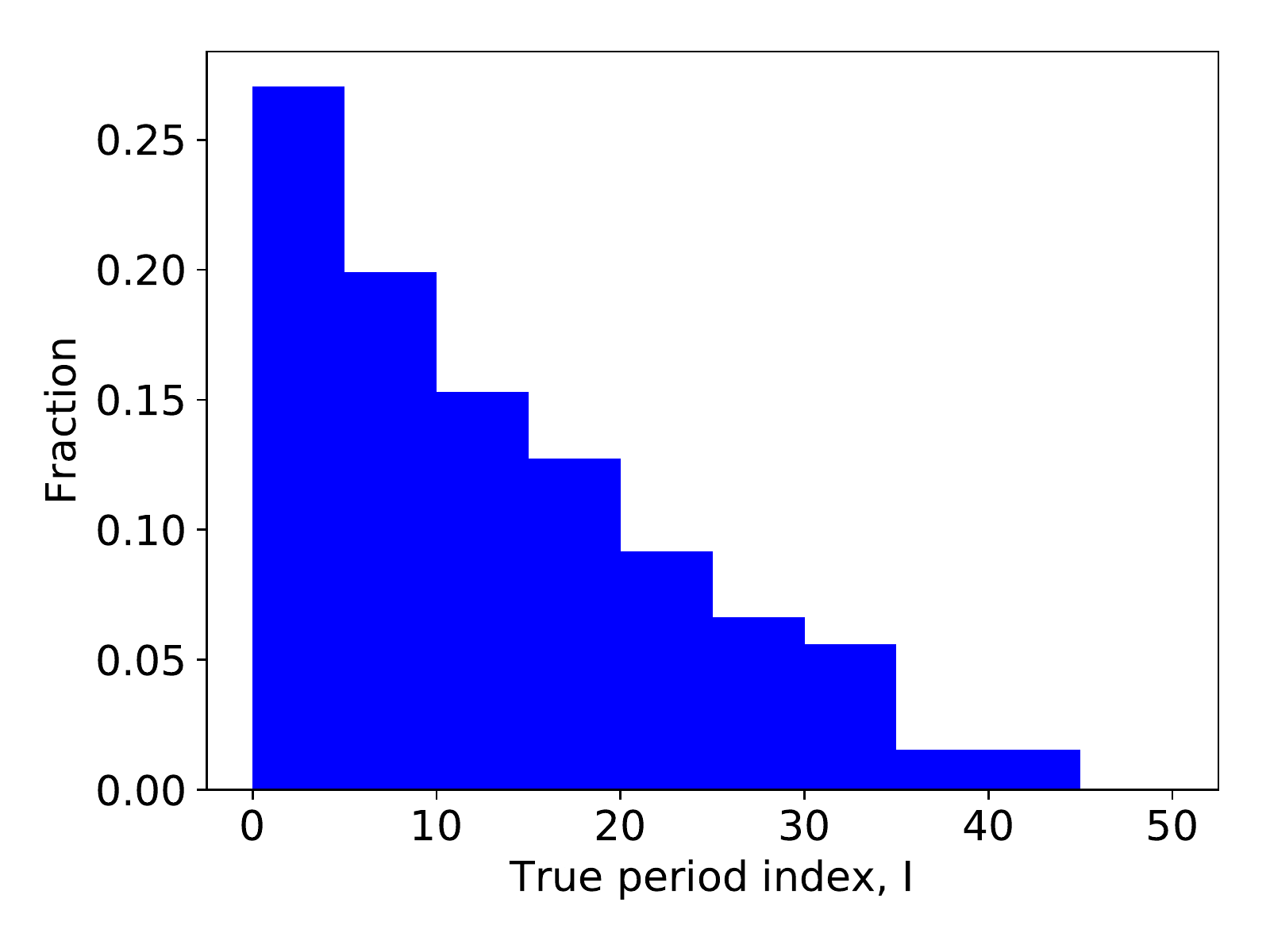}
\caption{Distribution of true period index, $I$. It can be seen that the distribution is maximal at $I = 0$ (i.e. shortest possible period is the true period) but there are a significant number of systems for which $I > 0$, and is non-negligible even up to large $I$ values.}
\label{fig:I}
\end{figure}

From these results we test whether the number and range of alias periods allows for the recovery of the true period. Based on the geometric probability of transit, 
one might assume that the true period is usually the shortest allowed period. However we find this is very often not the case.  In Figure \ref{fig:I} we show the distribution of the true period index $I$ which is the period index that corresponds to the true period.


\subsection{Extended mission timeline}
\label{sec:Extended mission timeline}

Since some of the specifics of the extended mission are still unknown we briefly discuss here the impact that may be expected if TESS were to return to the southern ecliptic hemisphere after two years instead of three years. First and foremost we would not expect a sizeable change in the number of recovered Year 1 monotransits. Each system would still be re-observed for the same amount of time so there is no reason that planets should exhibit significantly more or less additional transits after two or three years.

Part of this simulation that would be affected is the period alias discussion presented in Section~\ref{sec:Period alias}. If the southern hemisphere Sectors were re-observed after only two years the set of discrete possible periods for those Year 1 monotransits which show a second transit would be reduced. The average value of $T_{diff}$ would now be 2 years, or 730.5 days, a fraction of 2/3 as long as before. This then results in approximately 2/3 as many allowed periods for each system.

Some suggestions for the extended TESS mission include a number of TESS observing sectors placed along the ecliptic plane. Assuming these sectors to be exactly along the equator they will overlap the bottom $6^\circ$ of the southern hemisphere sectors. This will be $\frac{1}{16}^{\rm th}$ of the southern observations (if the whole ecliptic plane is observed). Without knowing when this is to occur it is hard to estimate its effect but qualitatively we would expect the recovered monotransits to increase by a similar fraction.


\section{Conclusions}
\label{sec:Conclusions}

Using the current understanding of the TESS stellar population, the planetary occurrence rates from Kepler, a realistic model of TESS photometric capabilities, and a realistic window function, we expect approximately 339 monotransit targets in the Year 1 data from TESS.  Of these targets, 266 will be seen to transit again in the Year 4 monitoring ($\sim80$\%), while 73 will remain as monotransit systems ($\sim20$\%). When this number is combined with our predictions of monotransits found in Year 4 only, we end the extended mission with 149 monotransiting systems with signals above the detection threshold.

The majority of Year 1 monotransits that are recovered in Year 4 will only show one transit in the Year 4 data (198/266, or approximately 75\%), leaving a wide range of potential periods.  Furthermore these are preferentially the longer period systems, which may be seen as more scientifically valuable as they cover planetary equilibrium temperatures that are interesting for internal structure studies \citep[e.g.][]{thorngren2018} or even for habitability studies \citep[e.g.][]{Schwieterman2019}.  To discover the true period either a spectroscopic orbit will be required, or the discrete set of periods can be used to predict transit times for photometric follow-up observations.  Which of these avenues is best will depend on a variety of factors including the magnitude of the host star, the transit depth, and the availability of spectroscopic or photometric resources (Cooke et al., in prep). Both techniques are being used successfully for recovery of TESS monotransit periods; for spectroscopy see Lendl et al. (in prep.) and for photometry see Gill et al. (in prep.). The number of monotransit candidates found so far in TESS is in broad agreement with the numbers presented here. In depth searches for these candidates is underway, as evidenced by these two in prep. papers.



\begin{acknowledgements}

BFC acknowledges a departmental scholarship from the University of Warwick.

\end{acknowledgements}

\bibliographystyle{aa}
\bibliography{letter.bib}

\begin{thebibliography}{20}
\expandafter\ifx\csname natexlab\endcsname\relax\def\natexlab#1{#1}\fi

\bibitem[{Akeson {et~al.}(2013)Akeson, Chen, Ciardi, {et~al.}}]{Akeson2013}
Akeson, R.~L., Chen, X., Ciardi, D., {et~al.} 2013, Publications of the
  Astronomical Society of the Pacific, 125, 989,
  \href{https://arxiv.org/pdf/1307.2944.pdf}{https://arxiv.org/pdf/1307.2944.pdf}

\bibitem[{Barclay {et~al.}(2018)Barclay, Pepper, \& Quintana}]{Barclay2018}
Barclay, T., Pepper, J., \& Quintana, E.~V. 2018, ArXiv e-prints, 1804.05050,
  \href{https://arxiv.org/pdf/1804.05050.pdf}{https://arxiv.org/pdf/1804.05050.pdf}

\bibitem[{Bouma {et~al.}(2017)Bouma, Winn, Kosiarek, {et~al.}}]{Bouma2017}
Bouma, L.~G., Winn, J.~N., Kosiarek, J., {et~al.} 2017, ArXiv e-prints,
  1705.08891,
  \href{https://arxiv.org/pdf/1705.08891.pdf}{https://arxiv.org/pdf/1705.08891.pdf}

\bibitem[{Cooke {et~al.}(2018)Cooke, Pollacco, West, {et~al.}}]{Cooke2018}
Cooke, B.~F., Pollacco, D., West, R., {et~al.} 2018, Astronomy and
  Astrophysics, 619, A175,
  \href{https://arxiv.org/pdf/1809.10687.pdf}{https://arxiv.org/pdf/1809.10687.pdf}

\bibitem[{Dressing \& Charbonneau(2015)}]{Dressing2015}
Dressing, C. \& Charbonneau, D. 2015, The Astrophysical Journal, 807, 45,
  \href{https://arxiv.org/pdf/1501.01623.pdf}{https://arxiv.org/pdf/1501.01623.pdf}

\bibitem[{Esposito {et~al.}(2019)Esposito, Armstrong, Gandolfi,
  {et~al.}}]{Esposito2019}
Esposito, M., Armstrong, D.~J., Gandolfi, D., {et~al.} 2019, Astronomy and
  Astrophysics, 623, A165,
  \href{https://arxiv.org/pdf/1812.05881.pdf}{https://arxiv.org/pdf/1812.05881.pdf}

\bibitem[{Fressin {et~al.}(2013)Fressin, Torres, Charbonneau,
  {et~al.}}]{Fressin2013}
Fressin, F., Torres, G., Charbonneau, D., {et~al.} 2013, The Astrophysical
  Journal, 766, 81,
  \href{https://arxiv.org/pdf/1301.0842.pdf}{https://arxiv.org/pdf/1301.0842.pdf}

\bibitem[{Gandolfi {et~al.}(2018)Gandolfi, Barrag{\'a}n, Livingston,
  {et~al.}}]{Gandolfi2018}
Gandolfi, D., Barrag{\'a}n, O., Livingston, J.~H., {et~al.} 2018, Astronomy and
  Astrophysics, 619, L10,
  \href{https://www.aanda.org/articles/aa/pdf/2018/11/aa34289-18.pdf}{https://www.aanda.org/articles/aa/pdf/2018/11/aa34289-18.pdf}

\bibitem[{{Huang} {et~al.}(2018){Huang}, {Burt}, {Vanderburg}, {G{\"u}nther},
  {Shporer}, {Dittmann}, {Winn}, {Wittenmyer}, {Sha}, {Kane}, {Ricker},
  {Vanderspek}, {Latham}, {Seager}, {Jenkins}, {Caldwell}, {Collins},
  {Guerrero}, {Smith}, {Quinn}, {Udry}, {Pepe}, {Bouchy}, {S{\'e}gransan},
  {Lovis}, {Ehrenreich}, {Marmier}, {Mayor}, {Wohler}, {Haworth}, {Morgan},
  {Fausnaugh}, {Ciardi}, {Christiansen}, {Charbonneau}, {Dragomir}, {Deming},
  {Glidden}, {Levine}, {McCullough}, {Yu}, {Narita}, {Nguyen}, {Morton},
  {Pepper}, {P{\'a}l}, {Rodriguez}, {Stassun}, {Torres}, {Sozzetti}, {Doty},
  {Christensen-Dalsgaard}, {Laughlin}, {Clampin}, {Bean}, {Buchhave}, {Bakos},
  {Sato}, {Ida}, {Kaltenegger}, {Palle}, {Sasselov}, {Butler}, {Lissauer},
  {Ge}, \& {Rinehart}}]{huang2018b}
{Huang}, C.~X., {Burt}, J., {Vanderburg}, A., {et~al.} 2018, \apjl, 868, L39,
  \href{https://arxiv.org/pdf/1809.05967.pdf}{https://arxiv.org/pdf/1809.05967.pdf}

\bibitem[{Huang {et~al.}(2018)Huang, Shporer, Dragomir, {et~al.}}]{Huang2018}
Huang, C.~X., Shporer, A., Dragomir, D., {et~al.} 2018, ArXiv e-prints,
  1807.11129,
  \href{https://arxiv.org/pdf/1807.11129.pdf}{https://arxiv.org/pdf/1807.11129.pdf}

\bibitem[{Jenkins {et~al.}(2016)Jenkins, Twicken, McCauliff,
  {et~al.}}]{Jenkins2016}
Jenkins, J.~M., Twicken, J.~D., McCauliff, S., {et~al.} 2016, in \procspie,
  Vol. 9913, Software and Cyberinfrastructure for Astronomy IV, 99133E,
  \href{https://www.spiedigitallibrary.org/conference-proceedings-of-spie/9913/1/The-TESS-science-processing-operations-center/10.1117/12.2233418.short?SSO=1}{https://www.spiedigitallibrary.org/conference-proceedings-of-spie/9913/1/The-TESS-science-processing-operations-center/10.1117/12.2233418.short?SSO=1}

\bibitem[{Ricker {et~al.}(2015)Ricker, Winn, Vanderspek, {et~al.}}]{Ricker2015}
Ricker, G.~R., Winn, J.~N., Vanderspek, R., {et~al.} 2015, Journal of
  Astronomical Telescopes, Instruments and Systems, 1, 014003,
  \href{https://arxiv.org/pdf/1406.0151.pdf}{https://arxiv.org/pdf/1406.0151.pdf}

\bibitem[{{Schwieterman} {et~al.}(2019){Schwieterman}, {Reinhard}, {Olson},
  {Harman}, \& {Lyons}}]{Schwieterman2019}
{Schwieterman}, E.~W., {Reinhard}, C.~T., {Olson}, S.~L., {Harman}, C.~E., \&
  {Lyons}, T.~W. 2019, \apj, 878, 19,
  \href{https://arxiv.org/pdf/1902.04720.pdf}{https://arxiv.org/pdf/1902.04720.pdf}

\bibitem[{Stassun {et~al.}(2019)Stassun, Oelkers, Paegert,
  {et~al.}}]{Stassun2019}
Stassun, K.~G., Oelkers, R.~J., Paegert, M., {et~al.} 2019, ArXiv e-prints,
  1905.10694,
  \href{https://arxiv.org/pdf/1905.10694.pdf}{https://arxiv.org/pdf/1905.10694.pdf}

\bibitem[{Stassun {et~al.}(2017)Stassun, Oelkers, Pepper,
  {et~al.}}]{Stassun2017}
Stassun, K.~G., Oelkers, R.~J., Pepper, J., {et~al.} 2017, ArXiv e-prints,
  1706.00495,
  \href{https://arxiv.org/pdf/1706.00495.pdf}{https://arxiv.org/pdf/1706.00495.pdf}

\bibitem[{{Thorngren} \& {Fortney}(2018)}]{thorngren2018}
{Thorngren}, D.~P. \& {Fortney}, J.~J. 2018, \aj, 155, 214,
  \href{https://iopscience.iop.org/article/10.3847/1538-3881/aaba13/pdf}{https://iopscience.iop.org/article/10.3847/1538-3881/aaba13/pdf}

\bibitem[{Vanderspek {et~al.}(2019)Vanderspek, Huang, Vanderburg,
  {et~al.}}]{Vanderspek2019}
Vanderspek, R., Huang, C.~X., Vanderburg, A., {et~al.} 2019, The Astrophysical
  Journal, 871, L24,
  \href{https://arxiv.org/pdf/1809.07242.pdf}{https://arxiv.org/pdf/1809.07242.pdf}

\bibitem[{Villanueva~Jr. {et~al.}(2018)Villanueva~Jr., Dragomir, \&
  Scott~Gaudi}]{Villanueva2018}
Villanueva~Jr., S., Dragomir, D., \& Scott~Gaudi, B. 2018, ArXiv e-prints,
  1805.00956,
  \href{https://arxiv.org/pdf/1805.00956.pdf}{https://arxiv.org/pdf/1805.00956.pdf}

\bibitem[{Wang {et~al.}(2019)Wang, Jones, Shporer, {et~al.}}]{Wang2019}
Wang, S., Jones, M., Shporer, A., {et~al.} 2019, The Astronomical Journal, 157,
  51,
  \href{https://arxiv.org/pdf/1810.02341.pdf}{https://arxiv.org/pdf/1810.02341.pdf}

\bibitem[{Winn(2014)}]{Winn2014}
Winn, J.~N. 2014, ArXiv e-prints, 1001.2010,
  \href{https://arxiv.org/pdf/1001.2010.pdf}{https://arxiv.org/pdf/1001.2010.pdf}

\end{thebibliography}

\begin{appendix}

\section{Year 1 updated results}
\label{sec:Year 1 updated results}

\begin{table*}[ht]
\centering
\caption{Numbers of simulated planets. This table displays planets by number of sectors that observe them as well as making splits based on detectability, single transit nature and observing cadence. Targets observed by 5-11 sectors are combined as individually they comprise very small fractions of the total. Also included is the percentage of sky area observed by each number of sectors and the percentage of total simulated planets in each region.}
\label{tab:numbers table}
\begin{tabular}{|c||c|c||c|c||c|c||c|c|}
\hline
{Number Of Sectors} & \multicolumn{2}{c||}{All Planets} & \multicolumn{2}{c||}{Detectable Planets} & \multicolumn{2}{c||}{Detectable Single Transits} & {Sky Area} & {Planets}\\ \cline{2-7}
Observed By                       & PS          & FFI          & \hspace{5 mm}PS\hspace{5 mm}              & FFI             &     \hspace{5 mm}PS\hspace{5 mm}                & FFI                &            (\%) & (\%)     \\ \hline\hline
0                &  0       &        779971        &  0         &      0            &   0           &    0      &   13.48    &         12.12  \\ \hline
1               & 45873     &      4080171        &         208   &          1572   &         37     &         246      &    64.08   &       64.14  \\ \hline
2               &  23675     &      1040112      &           150    &         542   &          12     &         36      &   15.08    &        16.54  \\ \hline
3               &  7289      &      187330      &    23     &         119         &    2              & 3      &    3.00    &         3.03   \\ \hline
4              & 2264        &    40121        &   11        &      30            &  0          &     0       &  0.67   &        0.66   \\ \hline
5 – 11           & 8262       &     83112     &      75  &            55  &            2     &          1   &    1.47    &   1.42   \\ \hline
12             & 289        &     1478       &     6           &    2            &   0           &    0          &   0.03   &     0.03   \\ \hline
13             & 21148      &     111866    &      142         &    117         &    0           &    0      &   2.20   &         2.07   \\ \hline\hline
Cadence Total          & 108800 & 6324161 & 615 & 2437 & 53 & 286 & 100.00 & 100.00  \\ \hline
Full Total                                & \multicolumn{2}{c||}{6432961}     & \multicolumn{2}{c||}{3052}              & \multicolumn{2}{c||}{339}                    &    \multicolumn{2}{c|}{}                       \\ \hline
\end{tabular}
\end{table*}

\begin{table}[ht]
\centering
\caption{Number of detectable planets found in each of 20 logarithmic period bins. Planets are separated into multitransiting planets and single transiting planets. Also shown is the number of sub-Neptune planets ($R_p \leq 4R_{\oplus}$). Planets with $P<1$ day are omitted.}
\label{tab:period table}
\resizebox{\columnwidth}{!}{\begin{tabular}{|c|c|c|}
\hline
Period Bin (days) & Multitransits ( $R_p \leq 4R_{\oplus}$ ) & Single Transits ( $R_p \leq 4R_{\oplus}$ ) \\ \hline\hline
1.0 - 1.4        &      88 ( 37 )        &      0 ( 0 ) \\ \hline
1.4 - 2.0        &      88 ( 49 )        &      0 ( 0 ) \\ \hline
2.0 - 2.8        &      274 ( 95 )       &      0 ( 0 ) \\ \hline
2.8 - 4.0        &      377 ( 103 )      &      0 ( 0 ) \\ \hline
4.0 - 5.6        &      432 ( 104 )      &      0 ( 0 ) \\ \hline
5.6 - 7.9        &      428 ( 160 )      &      0 ( 0 ) \\ \hline
7.9 - 11.2       &      349 ( 108 )      &      13 ( 1 ) \\ \hline
11.2 - 15.8      &      281 ( 85 )       &      28 ( 4 ) \\ \hline
15.8 - 22.4      &      150 ( 60 )       &      74 ( 11 ) \\ \hline
22.4 - 31.6      &      51 ( 23 )        &      95 ( 21 ) \\ \hline
31.6 - 44.7      &      36 ( 14 )        &      85 ( 15 ) \\ \hline
44.7 - 63.1      &      13 ( 4 )         &      22 ( 3 ) \\ \hline
63.1 - 89.1      &      4 ( 1 )          &      9 ( 2 ) \\ \hline
89.1 - 125.9     &      1 ( 0 )          &      8 ( 2 ) \\ \hline
125.9 - 177.8    &      1 ( 1 )          &      4 ( 0 ) \\ \hline
177.8 - 251.2    &      0 ( 0 )          &      1 ( 0 ) \\ \hline
251.2 - 354.8    &      0 ( 0 )          &      0 ( 0 ) \\ \hline
354.8 - 501.2    &      0 ( 0 )          &      0 ( 0 ) \\ \hline
501.2 - 707.9    &      0 ( 0 )          &      0 ( 0 ) \\ \hline
707.9 - 1000.0   &      0 ( 0 )          &      0 ( 0 ) \\ \hline
 \hline\hline
Total             & 2573 ( 844 )  & 339 ( 59 )      \\ \hline
\end{tabular}}
\end{table}

\end{appendix}




\end{document}